\documentclass[showpacs,preprintnumbers,amsmath,amssymb,twocolumn]{revtex4}
\usepackage{graphicx}% Include figure files
\usepackage{dcolumn}% Align table columns on decimal point
\usepackage{bm}% bold math

\newcommand{\be}{\begin{equation}}
\newcommand{\ee}{\end{equation}}
\newcommand{\lra}[1]{\langle #1 \rangle }

\begin{document}

\title{Extinction dynamics of a discrete population in an oasis}

\author{Stefano Berti}\affiliation{Laboratoire de M\'ecanique de
  Lille, CNRS/UMR 8107, Universit\'e Lille 1, 59650 Villeneuve d'Ascq,
  France} 

\author{Massimo Cencini}\affiliation{Istituto dei Sistemi Complessi,
  CNR, Via dei Taurini 19, 00185, Rome, Italy} 

\author{Davide Vergni}\thanks{Corresponding author}
\email{davide.vergni@cnr.it} \affiliation{Istituto per le Applicazioni
del Calcolo, CNR, Via dei Taurini 19, 00185, Rome, Italy} 

\author{Angelo Vulpiani}\affiliation{Dipartimento di Fisica,
  Universit\`a ``La Sapienza'',Piazzale Aldo Moro 2, I-00185 Roma,
  Italy}\affiliation{Istituto dei Sistemi Complessi,
  CNR, Via dei Taurini 19, 00185, Rome, Italy} 

\begin{abstract} 
Understanding the conditions ensuring the persistence of a population
is an issue of primary importance in population biology.  The first
theoretical approach to the problem dates back to the 50's with the
KiSS (after Kierstead, Slobodkin and Skellam) model, namely a
continuous reaction-diffusion equation for a population growing on a
patch of finite size $L$ surrounded by a deadly environment with
infinite mortality -- i.e. an oasis in a desert. The main outcome of
the model is that only patches above a critical size allow for
population persistence.  Here, we introduce an individual-based
analogue of the KiSS model to investigate the effects of discreteness
and demographic stochasticity. In particular, we study the average
time to extinction both above and below the critical patch size of the
continuous model and investigate the quasi-stationary distribution of
the number of individuals for patch sizes above the critical
threshold.
\end{abstract}

\maketitle

\section{Introduction}
Many biological or chemical processes involve the dynamics of discrete
``particles'' (e.g., molecules or organisms) that diffuse and interact
with each other and/or with an external
environment~\cite{murray1993,fglo1999,dewitt1999,tk2001,tdgk2005,
fedotov2008,neufeld2009}.  In
population dynamics, for instance, individuals spread in space,
interact, reproduce and die.  

If the number density of particles is very large, the macroscopic
description in terms of continuous fields is typically appropriate.  A
well established approach to model the spatio-temporal evolution of the
population density field, $\mathcal{P}(x,t)$, is in terms of a
reaction-diffusion equation that, in one spatial dimension, reads
\begin{equation}
\frac{\partial \mathcal{P}}{\partial t} = D \frac{\partial^2
  \mathcal{P}}{\partial x^2} + f(\mathcal{P})\,.
\label{eq:rd}
\end{equation}
In the above equation, $D$ is the diffusion coefficient and
$f(\mathcal{P})$ rules the chemical kinetics or, 
in the language of population biology, the local rate of population
growth. A  well known growth term is the logistic model 
\begin{equation}
f(\mathcal{P}) = r \mathcal{P} \left(1 - \frac{\mathcal{P}}{K}\right)\,,
\label{eq:logistic}
\end{equation}
whose quadratic term models the competition for resources,
$K$ being the local carrying capacity (i.e. the maximal population
density that can be sustained) while $r$ denotes the intrinsic growth
rate.

Conversely, if the density of particles is not very large the discrete
nature of the population cannot be neglected~\cite{dl1994}, new
effects arise and the continuous description becomes inaccurate.  In
order to account for the new effects several approaches are possible.
For instance, for a population of $N$ individuals, a possibility is to
introduce a cutoff at the value $1/N$ of the normalized density for
the continuous field equations~\cite{bd1997} (see \cite{panja2004} for
a review).  Another possibility is to supplement Eq.~(\ref{eq:rd})
with a noise term accounting for microscopic fluctuations originated
by the finite number of
individuals~\cite{levine1999,dms2003,hallatschek2009}.  A further
modeling step is to consider a lattice where each site is occupied by
an integer number of individuals~\cite{Geyrhofer2013,Grassberger2013},
or a contact process (see, e.g.,~\cite{dl1994b,jl2005} and references
therein), where each site is occupied at most by an individual.

In the present work, we consider a system of particles diffusing in
space and interacting when they get within a given interaction
distance. In particular, revisiting the continuous KiSS model, after
Kierstead, Slobodkin and Skellam \cite{skellam1951,ks1953}, we
investigate the effects induced by discreteness on the dynamics of a
population inhabiting a favorable region (``an oasis'') surrounded by
a deadly environment (``a desert'').  Accounting for such effects is
important to assess the role of demographic and environmental
stochasticity on extinction dynamics~\cite{lande1993}.

 Several efforts have already been undertaken along this direction
 within the framework of the KiSS model. Some model modifications that
 include effects of intrinsic variability on the critical habitat size
 ensuring population persistence have been studied in relation to
 environmental stochasticity \cite{Mendez2010}. As for demographic
 stochasticity due to population discreteness it has been investigated
 either using a cutoff on the continuous field equations
 \cite{Mendezcutoff}, in the same spirit of Ref.~\cite{bd1997}, or
 within a stochastic framework using field theory
 techniques~\cite{escudero2004}.  In our work, we adopt a discrete
 particle model \cite{blvv2007} focusing on demographic stochasticity
 effects. Moreover, our study is not restricted to steady state
 properties but also deals with dynamical features of extinction
 phenomena.

The article is organized as follows. In Section~\ref{sec:surv} we
recall the continuous KiSS model~\cite{skellam1951,ks1953} and
introduce its discrete analogue. In Section~\ref{sec:qss} we
numerically study the long time properties of the discrete model,
characterizing the so-called quasi-stationary state
\cite{ovaskainen2001,dv2002}, and comparing it with the properties of
the continuous model.  Section~\ref{sec:mte} is devoted to the problem
of determining the critical habitat size for population persistence
and to the estimation of extinction times in the discrete particle
system.  Discussions and conclusions are presented in
Sec.~\ref{sec:concl}. As a side discussion, in the Appendix we briefly
consider the effects of changes of the microscopic rules of the
discrete model.

\section{Model}
\label{sec:surv}

\subsection{Continuous KiSS model}
\label{sec:kiss_cont}
Determining the conditions for persistence or, conversely, estimating
the probability of extinction of species and populations is a question
of paramount importance in population biology.  The theoretical
treatment of the problem was initiated by Kierstead, Skellam and Slobodkin
\cite{skellam1951,ks1953} who considered a population
that, ruled by Eqs.~(\ref{eq:rd}-\ref{eq:logistic}), grows with rate
$r$ and diffuses with diffusion coefficient $D$ within a patch of size
$L$ surrounded by a deadly environment. For a given growth rate,
higher diffusivities imply larger fluxes across the boundaries, so
that larger patches are necessary to compensate the population loss in
order to allow stable persistence.  It is then natural to
determine the minimal (critical) patch size,
$L_c$, ensuring population persistence.  At least dimensionally, the
critical patch size can be obtained by balancing the growth rate $r$
with the exit rate from the patch due to diffusion $D/L^2$
\cite{ol2001}. This way one derives that $L_c \sim
\sqrt{D/r}$ that is the correct result up to an order one constant 
(see below).

In one-dimensional space, for simplicity, it is enough to consider
Eqs.~(\ref{eq:rd}-\ref{eq:logistic}) in the interval $0\leq x \leq L$,
representing the favorable patch, while the outside hostile
environment results in the boundary conditions
\begin{equation}
\mathcal{P}(0) = \mathcal{P}(L) = 0\,.
\label{eq:ckiss_bc}
\end{equation}

Detailed analysis of the KiSS model can be found in
\cite{skellam1951,ks1953,ol2001,rb2008,mendezbook}, for
generalizations of the model including different kind of population
growth terms see \cite{Mendezallee1,Mendezallee2}. Here, we only
recall the main results on the critical patch size. We notice that,
close to the extinction, the population density will be very small
($\mathcal{P}\ll K$) so that we can linearize the growth term
(\ref{eq:logistic}) by posing $f(\mathcal{P})=r\mathcal{P}$.  Starting
from a generic initial condition, population density will thus grow
(decay) depending on the positive (negative) sign of the largest
eigenvalue $\lambda_1$ of the (linear) operator $\mathcal{L} = D
\partial_x^2 + r$ with boundary conditions~(\ref{eq:ckiss_bc}).
Standard computation shows that the eigenvalues are
\begin{equation} 
\lambda_n = r \left[
  1 - n^2 \left(\frac{L_c}{L}\right)^2 \right], \quad n=1,2,...
\label{eq:ckiss_ev}
\end{equation}
with $L_c=\pi \sqrt{D/r}$. Consequently, the population can persist
only if the favorable habitat is larger than the critical patch size
$L_c$.   However, while in the linear case also the
eigenfunctions and thus the stationary population density can be
easily derived, for the logistic case only approximate
solutions are known (see, e.g., Ref. \cite{skellam1951}).

It is worth remarking that proper estimations of $L_c$, with suitable
generalizations of the above model, are used in the design of
protected areas for endangered species, in the context of conservation
biology~\cite{dm1976,cc1998,cantrell1999}.

In the following we focus on the logistic growth case. For later
convenience it is useful to formulate the continuous model
(\ref{eq:rd}-\ref{eq:logistic}) in non-dimensional variables.  From
the above analysis it is natural to measure time in units of the
inverse growth rate $1/r$ and space in units of $\sqrt{D/r}$. Finally,
normalizing the population density to the carrying capacity, that is
introducing $\theta=\mathcal{P}/K$, we can rewrite
Eqs.~(\ref{eq:rd}-\ref{eq:logistic}) as
\begin{equation}
\frac{\partial \theta}{\partial t} = \frac{\partial^2
  \mathcal{\theta}}{\partial x^2} + \theta(1-\theta)\,.
\label{eq:ckiss_lo_nd}
\end{equation}
with $\theta(0)=\theta(L)=0$, where $L$ now is 
the non-dimensional domain size. With these units the critical patch size is
$L_c = \pi$.

\subsection{Discrete KiSS model}
\label{sec:kiss_disc}
In recent years, several efforts have been devoted to the extinction
problem in populations with a finite number, $N$, of individuals,
using non-spatial models~\cite{lande1993,nasell2001,dss2005,om2010}.
The factors influencing extinction, through fluctuations and decline
of a population, are in principle extremely varied and attributable to
a wide class of biological and environmental causes, among which
demographic stochasticity~\cite{om2010}, associated to the unavoidable
random occurrence of birth and death events, is one of the most
important ones.

In the context of spatially extended systems, some effects of
demographic stochasticity on steady patterns of populations have been
analyzed using different models~\cite{escudero2004,Mendezcutoff}.  In
particular, in Ref.~\cite{Mendezcutoff} the discreteness was modeled
using reaction-diffusion equations with a cutoff on the growth term,
and it was found that the critical patch size for survival is larger
than in the model without cutoff.  A similar effect was found also in
Ref.~\cite{escudero2004}, using a stochastic discrete model, where it
is shown that the population can go extinct above the critical patch
size of the continuous model and an ``effective'' (finite) critical
patch size can be introduced only when the competition term is weak
enough.  In our study, similarly to \cite{escudero2004,Mendezcutoff},
the discrete population can go extinct for habitat sizes guaranteeing
the persistence in the continuous case. However, we will always refer
to the critical patch size of the continuous model $L_c$ because, as
we will see, this value still marks a transition in the system
behavior.

We now introduce a discrete version of the KiSS model,
  generalizing the stochastic particle model of Ref.~\cite{blvv2007}.
  As an essential  prerequisite, when the number of particles is large,
  the model must reproduce the FKPP (Fisher-Kolmogorov-Petrovskii-Piskunov) 
  dynamics $\partial_t \theta = D \partial^2_x \theta + r\theta(1-\theta)$ 
  (i.e., Eq.~(\ref{eq:ckiss_lo_nd}) in the dimensional version).  
  The FKPP equation can be derived from two coupled reaction-diffusion 
  equations for the concentrations $\theta_A$ and $\theta_B$ of the species 
  undergoing the autocatalytic reaction
\begin{equation}
 A + B \stackrel{r}{\longrightarrow} \,\, 2B\,.
   \label{eq:autocatalytic}
\end{equation}
Summing the dynamical equation for $\theta_A$ ($\partial_t \theta_A =
D_A \partial^2_x \theta_A - r\theta_A\theta_B$) with that for
$\theta_B$ ($\partial_t \theta_B = D_B \partial^2_x \theta_B +
r\theta_B\theta_B$) gives the FKPP equation, provided $D_A=D_B=D$ and
$\theta_A+\theta_B=1$ (i.e. local mass conservation).  We discretize
these reaction-diffusion equations passing from the concentrations
$\theta_{A,B}$ to a particle description. Thus, we consider $N_A$
particles of type $A$ and $N_B$ of type $B$, with $N=N_A+N_B$
fixed. All the particles undergo independent diffusive processes with
the same diffusion coefficient, $D$, within the favorable  patch
  $[0,L]$. Their positions $x_\alpha$ ($\alpha=1,\ldots,N$) diffuse
with diffusion coefficient $D$.  As for the interaction between
particles, a simple way to implement the interaction term
$r\theta_A\theta_B$ is to impose that a particle of type $A$ changes
into $B$ with a probability rate $W_{AB}$ depending on the intrinsic
rate, $r$, and on the number of $B$ particles within an interaction
distance $R$. This is mathematically formalized by the expression
\begin{equation}
   W_{AB} = r  \frac{n_B(x_\alpha;R)}{n_{av}(R)}
        = r \frac{n_B(x_\alpha;R)}{2 R \rho},
   \label{eq:probreaction}
\end{equation}
where $n_B(x_\alpha;R)$ denotes the number of $B$-particles within a
distance $R$ from the given $A$-particle located at $x_\alpha$, and
$n_{av}(R)=2 R \rho$ the average number of particles (regardless their
type) within the interaction interval of length $2R$ ($\rho = N/L$
being the particle density).  As discussed in Ref. \cite{blvv2007},
whenever the local particle density is large enough, the probabilistic
rule above described converges to the FKPP equation. Finally, we have
to fix the behavior of $A$- and $B$- particles at the boundaries.  In
order to reproduce the boundary conditions of the KiSS model and
locally ensuring mass conservation, we  impose the following
rules: $A$ particles hitting the boundaries are reflected (nutrients
are present only within the patch), if a $B$ particle hits the
boundary it is absorbed (as it cannot survive outside the favorable
patch) and  it is replaced in-place by an $A$-particle. 

The $A$ particles are used as ``auxiliary'' particles in order to
recover as a limit the continuous FKPP model. From an ecological
perspective, considering the $A$ particles as nutrient, it would be
interesting to study the case in which $A$ and $B$ particles diffuse
with different diffusion coefficients. Clearly, in this case the FKPP
cannot be expected to be the proper continuous description. Here we
limit our discussion to the case of same diffusion coefficients for 
the two species of particles. In the Appendinx, however, as an example we
briefly consider the case of non-diffusing $A$ particles.

From the algorithmic point of view, the above model consists of two
basic steps: the diffusive step -- in which all particle positions
$x_{\alpha}$ ($\alpha=1,\ldots,N$) evolve according to the dynamics
$dx_\alpha/dt=\sqrt{2D}\,\eta_{\alpha}$, with $\{\eta_{\alpha}\}$
independent normal random variables; and the reaction step -- in
which, for each $A$-particle, one counts all the $B$-particles within
a distance $R$ and changes $A$ in $B$ with probability $W_{AB}\Delta
t$. Whenever the time step $\Delta t$ is small enough the order in
which the two steps are performed is irrelevant.

The main difference between the continuous model and the discrete one
is that while the (continuous) stationary state $\theta=0$ becomes
unstable above the critical patch size, the (discrete) absorbing state
$N_B=0$ can always be reached due to demographic stochasticity. In the
following sections we will investigate in depth this issue.  Here, we
simply observe that the presence of an absorbing state forces us to
perform ensemble averages in numerical simulations. Therefore, our
simulations consist of a large number, $n_r$, of statistically
independent realizations. As for initial conditions, each realization
starts with $N/2$ particles of each type uniformly distributed within
the patch. Each realization is then followed until the absorbing state
($N_B=0$) is reached.  The interaction distance, $R$, is chosen to be
small compared to the size of the favorable patch but large enough to
contain some particles on average ($R\rho>1$), which provides a lower
bound on the minimum number of particles that one can consider,
typically $N \gtrsim 20$.  In all simulations we set $R=0.1$, $D=1$
and $r=1$ so that we have as reference in the continuum limit the
non-dimensional equation~(\ref{eq:ckiss_lo_nd}). Finally, the time
step, $\Delta t$, is chosen small enough such that the diffusive step
is much smaller that the interaction distance (i.e. $\sqrt{2D\,\Delta
  t}\ll R$). Here, we have chosen $\Delta t=10^{-4}$.

 We notice that the model is robust with respect to small changes of
 $R$, while if $R$ becomes comparable with the patch size relevant
 differences arise, as discussed in the Appendix.  There, we also
 briefly consider the subtle effects caused by modifications to the
 rules specifying the dynamics of the individual-based model.

\begin{figure}[ht!]
\centering
\includegraphics[scale=0.7]{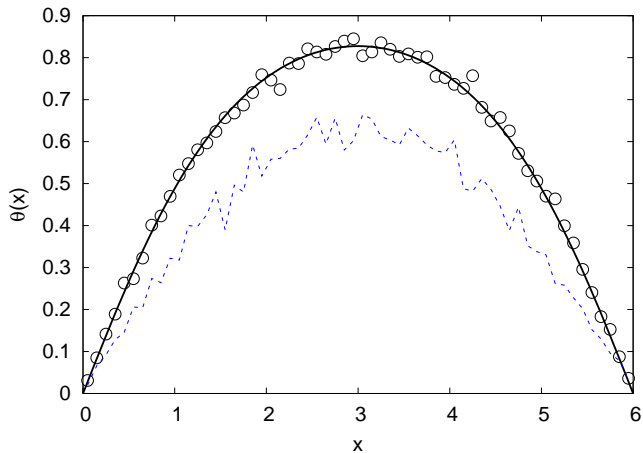}
\caption{(Color online) Density field $\theta(x)$ at
    stationarity for the continuous KiSS model (black solid curve)
    compared with the local density of $B$-particles, averaged over
    $n_r=10$ realizations, in the discrete model (empty circles).  The
    main parameters are: $N=10^3$, $R=0.1$, $L=6$.
    The (blue) dashed curve represents the local
    density of $B$-particles obtained with a
    different model with fixed $A$
    particles (see Appendix for details).}
\label{fig:1}
\end{figure}

As an example, we show in Fig.~\ref{fig:1} the good agreement, at
least for large $L$, between the stationary
population density, $\theta(x, t\to \infty)$, of the continuous KiSS
model and the long time (but finite, see Sect.~\ref{sec:qss}) particle
density, obtained by ensemble averaging the spatial distribution of
$B$ particles in the discrete model.

\section{The quasi-stationary state\label{sec:qss}}

In the discrete model, the presence of an absorbing state ($N_B=0$)
makes population extinction possible for any finite $N$, even when
$L>L_c$. Nevertheless, the time taken by the system to be absorbed can
be very long (for $N$ and $L$ large enough) and, in that case, the 
system reaches a quasi-stationary state.
The quasi-stationary distribution is defined as the probability
distribution conditioned on non-extinction
\cite{nasell1999,nasell2001,ovaskainen2001,dv2002,gw2004,dss2005}.
In general, it provides
interesting information about the properties of systems with an
absorbing state \cite{ovaskainen2001,dv2002} but, except in some rare
cases, it cannot be evaluated explicitly.
\begin{figure}[!htb]
\centering
\includegraphics[scale=0.6]{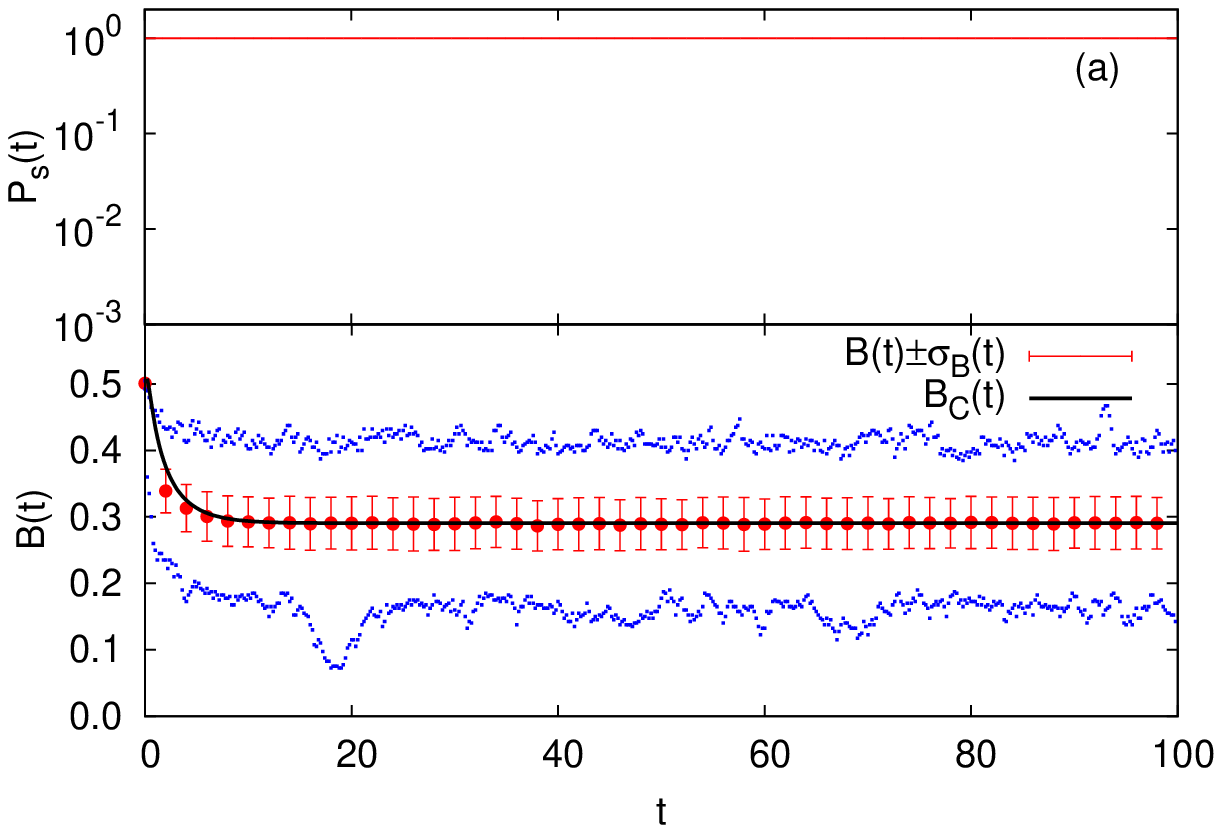}
\includegraphics[scale=0.6]{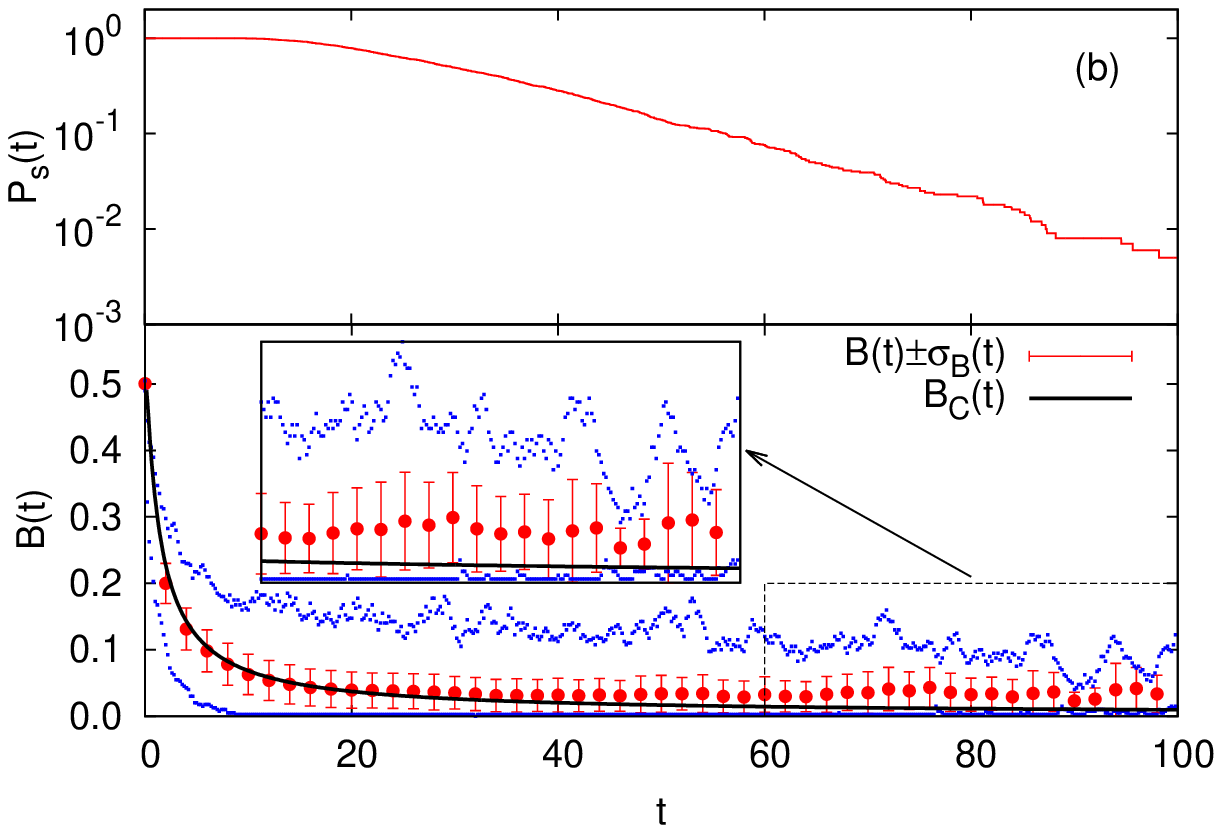}
\caption{(Color online) Survival probability and biomass as a function
  of time for two values of the patch size above the critical value
  $L_c=\pi$: $L=4$ (a) and $L=3.15$ (b), with $N=400$ particles.  Top
  panels show the survival probability $P_s(t)$, see text. Bottom
  panels show the time evolution of the continuous
  ($\mathcal{B}_C(t)$) and discrete ($\mathcal{B}(t)$) biomasses, the
  error bars on the latter display the standard deviation,
  $\sigma_\mathcal{B}(t)$.  The (blue) dotted curves denote the
  minimum and maximum values of $N_B(t)/N$ observed in all $n_r=1000$
  realizations.  The inset in (b) zooms the framed area in the main
  figure to better show that $\mathcal{B}(t)>\mathcal{B}_C^\infty$ at
  long times.
\label{fig:2}}
\end{figure}

As shown in Fig.~\ref{fig:1}, the quasi-stationary distribution, apart
from fluctuations, essentially reproduces the results obtained with 
the continuous model, at least when $L$
and $N$ are sufficiently large. It is then natural to study under
which conditions this correspondence holds. In particular, in
this section, we focus on the behavior of the biomass per unit length
defined as (see Refs.~\cite{young2007,vibc2012})
\begin{equation}
\mathcal{B}_C(t) = \frac{1}{L} \int_0^{L} \!\theta(x,t) dx\,.
\label{eq:biom_cont}
\end{equation}
The label $C$ simply recalls that this quantity is
defined in the continuum limit, and  allows to distinguish it from the
analogous quantity in the discrete case, for which no labels will be
used.  We observe that $\mathcal{B}_C(t)$ asymptotically vanishes when
$L<L_c$, due to population extinction, and approaches a positive steady
value, indicating survival, when $L>L_c$.  The asymptotic
value $\mathcal{B}^\infty_C=\lim_{t \to \infty} \mathcal{B}_C(t)$ can
be used as an order parameter to define the extinction/survival
transition.

In the discrete model, the biomass per unit length
(\ref{eq:biom_cont}) is nothing but the ratio between the total number
of $B$-particles and the total number of particles, $N_B(t)/N$. Since
$N_B$ is a time-fluctuating quantity, we consider its ensemble average
\begin{equation}
\mathcal{B}(t) = \left\langle \frac{N_B(t)}{N} \right\rangle\,, 
\label{eq:biom_disc}
\end{equation}
where the brackets $\langle \cdots \rangle$ denote averaging over
different realizations with the same initial conditions and
conditioning on survival (i.e. at each time $t$ only the realizations
with $N_B(t) \geq 1$ are considered).  When conditioning on
non-extinction, at long times, $\mathcal{B}(t)$ approaches a
definite value (Fig.~\ref{fig:2}) which is the average over the
quasi-stationary distribution.

In the bottom panels of Figure~\ref{fig:2}a,b we compare the time
evolution of $\mathcal{B}_C(t)$ and $\mathcal{B}(t)$ for two different
values of $L>L_c$. For the discrete case we also show, as error bars,
the standard deviation of the ratio $N_B(t)/N$,
\begin{equation}
\sigma_\mathcal{B}(t)=\left\langle
(N_B(t)/N-\mathcal{B}(t))^2\right\rangle^{1/2}\,.
\label{eq:Bfluct}
\end{equation}
  In the top panels, we
show the probability, $P_s(t)$ (measured as the fraction of
realizations such that $N_B(t)\geq 1$ at time $t$), that the
population survived up to time $t$.

For any given $N$, when $L$ is large enough (Fig.~\ref{fig:2}a),
$\mathcal{B}(t)$ essentially coincides with the continuum-limit and
all realizations survive ($P_s(t)=1$ in the observation window) for
long time.  In these circumstances, the minimal values of
$\mathcal{B}(t)$ (lower blue dotted curve in Fig.~\ref{fig:2}a) are
well above zero, implying that very unlikely fluctuations are needed
to reach the absorbing state.  Conversely, as $L$ approaches the
critical value $L_c$ (Fig.~\ref{fig:2}b) this is no longer true. As
shown in the top panel of Fig.~\ref{fig:2}b, $P_s(t)$ exponentially
decays to zero at a relatively short time.  More importantly, the
agreement between $\mathcal{B}(t)$ and the continuum limiting value
gets poorer.  In particular, at long times, $\mathcal{B}(t)$
approaches a quasi-stationary value which is larger than the
corresponding continuum one  (Fig.~\ref{fig:2}b, inset). This is
clearly a consequence of conditioning on non-extinction which leads to
an overestimation of the biomass, for these values of the patch size.

\begin{figure}[!t]
\centering
\includegraphics[scale=0.7]{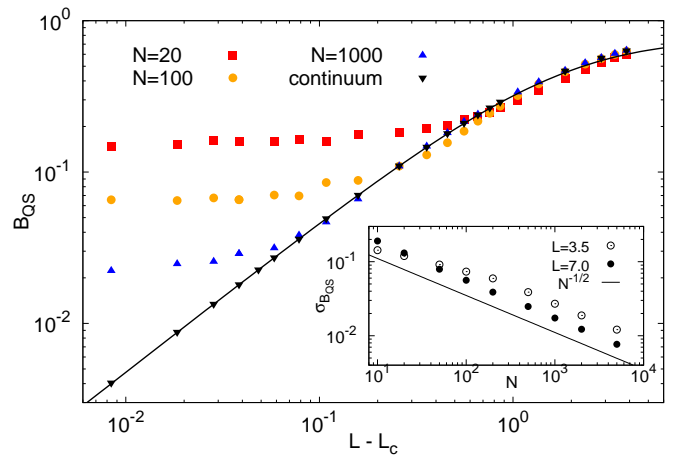}
\caption{(Color online) Limiting biomass $B_C^\infty$ and average
  biomass in the quasi-stationary state, $\mathcal{B}_{QS}$, as a
  function of $L-L_c$ for different values of $N$ as in legend.  
  The solid curve displays $0.75 \lambda_1$ that corresponds to the
  behaviour Eq.~(\ref{eq:biom_L}). The factor 0.75 cannot be catched
  by the argument exposed in the main text.  Inset: standard
  deviation of the biomass (\ref{eq:Bfluct}) in the quasi-stationary
  regime, $\sigma_{\mathcal{B}_{QS}}$, as a function of $N$ for two
  values ($L=3.5$ and $L=7$) of the patch size.  The number of
  realizations used for the averages ranges from $500$ to $10000$
  depending on the value of $N$. \label{fig:3}}
\end{figure}

The latter observation becomes more quantitative when looking at
Fig.~\ref{fig:3} where we compare, as a function of $L-L_c$, the
limiting value $\mathcal{B}^\infty_C$ with the quasi-stationary mean
value $\mathcal{B}_{QS}$, estimated by averaging $\mathcal{B}(t)$ over
the time interval in which it fluctuates around a constant value (see
Fig.~\ref{fig:2}).  We have also computed $\mathcal{B}_{QS}$ using the
algorithm proposed in Ref.~\cite{Dickman_sim2005}, which directly
probes the quasi-stationary distribution, obtaining (not shown)
indistinguishable results.  The plateau $\mathcal{B}_{QS}\approx const
>\mathcal{B}^\infty_C$, observed when $L\to L_c$, is essentially due
to conditioning on non-extinction. The effect is stronger the smaller
is the number of particles.  
The dependence of $\mathcal{B}_{QS}$
on the number of particles $N$ clearly demonstrates the departure from
the continuum limit when $L$ approaches the critical value. However,
for any $L>L_c$ there exists a value of $N$ (which diverges as $L\to
L_c$) above which $\mathcal{B}_{QS}$ tends to the continuous value
$\mathcal{B}_{C}^\infty$.

When the population survive, it is possibile to show, by using a
meanfield-like argument, that the biomass of the continuous field can
be approximated by
\begin{equation}
\mathcal{B}_C^\infty(L) \approx \lambda_1\,,
\label{eq:biom_L}
\end{equation}
as confirmed in Fig.~\ref{fig:3}.  The idea is to approximate the
right hand side of Eq.~(\ref{eq:ckiss_lo_nd}) as $\partial^2_{x}\theta
+\theta(1-\overline\theta)$, where $\overline\theta$ denotes the
steady state spatially averaged value. The first eigenvalue of this
problem, $\lambda^*_1$, is linked to the eigenvalue of the linear KiSS
model~(\ref{eq:ckiss_ev}) via $\lambda^*_1=\lambda_1-\overline\theta$.
To have a stationary solution one has to impose $\lambda^*_1=0$, so
that $\overline\theta=\lambda_1$. But at stationarity
$\overline\theta=\mathcal{B}^\infty_C$, which implies
~(\ref{eq:biom_L}). Of course this kind of argument cannot be expected
to work when $\theta$ is close to 1. It is worth observing that, for
$L-L_c\ll 1$, Eq.~(\ref{eq:biom_L}) implies that the continuous
biomass is linear in $L-L_c$.

\begin{figure}[!t]
\centering
\includegraphics[scale=0.7]{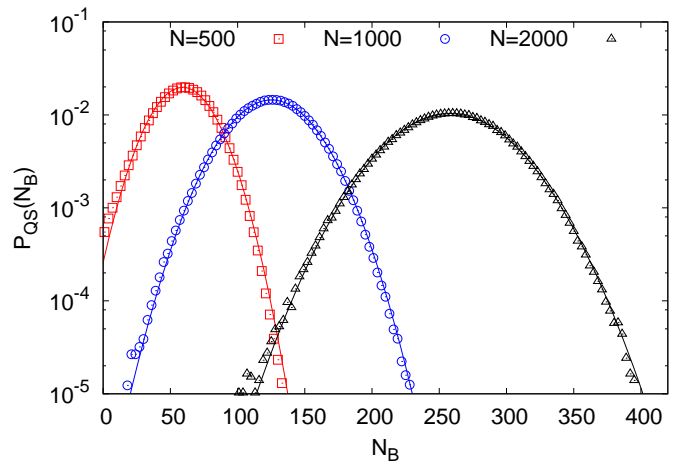}
\caption{(Color online) Quasi-stationary distribution
    for $L-L_C=0.3$ and three values of $N$ as in label. The solid
    curves display the Gaussian distribution obtained using the mean
    and standard deviation values measured from data. We observe a
    fairly good agreement at least for $N$ large enough. When $L\to
    L_C$ and/or $N$ is not large enough the agreement ceases to be so
    good, as one can observe from the left tail at
    $N=500$. \label{fig:4}}
\end{figure}

The inset of Fig.~\ref{fig:3} shows the standard deviation of the
biomass in the quasi-stationary state, $\sigma_{\mathcal{B}_{QS}}$, at
varying $N$, for two values of $L$.  At least for $L$ and $N$ large
enough, the numerical results indicate that $\sigma_{\mathcal{B}_{QS}}
\sim N^{-1/2}$.  We notice that the inverse square-root dependence of
$\sigma_{\mathcal{B}_{QS}}$ on $N$ implies for the standard deviation
of the number of $B$-particles $\sigma_N=N\sigma_{\mathcal{B}_{QS}}
\sim N^{1/2}$, suggesting that the distribution of $N_B$ particles is
Gaussian, as confirmed in Fig.~\ref{fig:4}.  This observation will be
exploited in the next section.

\section{Critical patch size and extinction times in the discrete model}
\label{sec:mte}
In the continuous KiSS model, the critical patch size $L_c$
discriminates between populations able to persist ($L>L_c$) and those
doomed to extinction ($L<L_c$).  For its ecological importance, many
efforts have been devoted to understanding how this critical length
is modified in less idealized situations, e.g. taking into
account more complex heterogeneous
environments~\cite{skt1986,kkts2003} and/or the presence of external
factors like fluid advection~\cite{abraham1998,dahmen2000life,rb2008,vibc2012}.  On
the other hand, as previously mentioned, in the case of the
individual-based model extinction can happen even when $L>L_c$.  It is
then natural to wonder how the survival/extinction transition of the
continuous KiSS model translates in the behavior of the average time
to extinction $T_e$ \cite{om2010}, which is linked to 
the survival probability $P_s(t)$ as $T_e = \int_0^\infty
\!\!P_s(t)\,{\mathrm d}t$~\cite{redner2001}. Computations of mean extinction
times have been mainly carried on in non-spatial models, where the
dependence of the extinction time $T_e$ on the number of individuals
$N$ at varying the population growth rate has been obtained using
diffusive approximations \cite{lande1993}.  These results have been
influential for their rather simple mathematical formulation and
generality.  However, it has been recently recognized that the
diffusive approximation can give wrong answers
\cite{nasell2001,ovaskainen2001,dss2005}.

\begin{figure}[!b]
\centering
\includegraphics[scale=0.7]{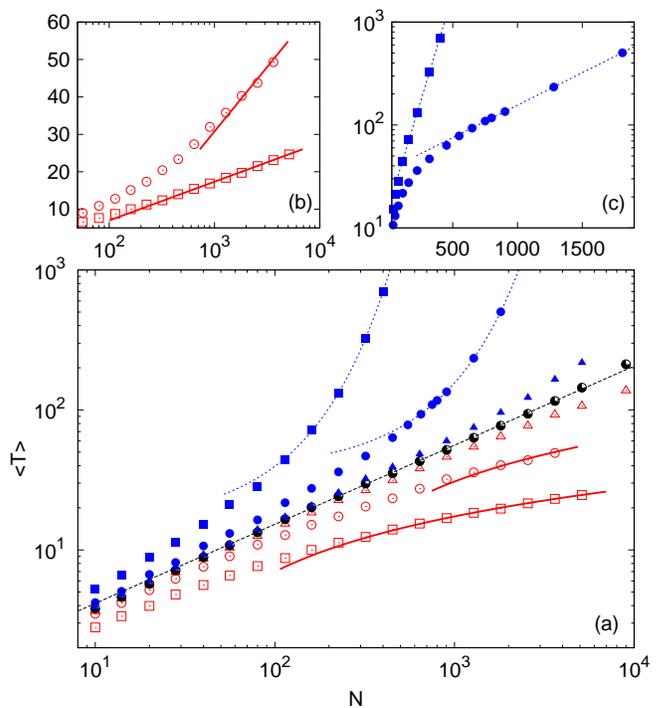}
\caption{(Color online) Mean extinction time, $T_e$, vs population
  size, $N$, for different patch sizes $L=L_c +\delta L$. Panel (a)
  from bottom to top $\delta L= -0.3, -0.1,-0.02,0,0.02,0.1,0.3$,
  encompassing the extinction (empty red symbols) and persistence
  (blue filled symbols) regions, as well as the critical point (black
  half filled symbols) of the continuous model.  The continuous red
  lines display the prediction (\ref{eq:Te_ext}) with $b'$ fitted from
  data; the dashed blue lines correspond to the exponential behavior
  (\ref{eq:Te_pers}) with $a$ fitted. For $L=L_c$ we found a power-law
  behavior, $T_e \sim N^{\gamma}$, with best fitting exponent
  $\gamma=0.565$ (see text).  Panels (b) and (c) display a zoom of the
  main figure, in appropriate semilog scale, of both the logarithmic
  (\ref{eq:Te_ext}) and the exponential (\ref{eq:Te_pers}) regime,
  respectively. We emphasize that the slope in (b) is not fitted but
  obtained from the first eigenvalue. Depending on the value of $N$
  the number of realizations ranges from $500$ to
  $5000$.\label{fig:5}}
\end{figure}

In what follows we examine the dependence of $T_e$ on the population
size $N$ in the extinction ($L<L_c$), critical ($L=L_c$) and
persistent ($L>L_c$) regions of the continuous KiSS model.  The
different regions, indeed, are characterized by unalike functional
dependencies of the time to extinction $T_e$ on $N$ as shown in
Fig.~\ref{fig:5}. We observe that quantitatively $T_e$ depends on the
initial condition, however the dependence on $N$ and $L$, which is
here explored, is qualitatively independent of the initial condition,
provided the initial population is not too small in order to avoid
spurious fast extinctions. Therefore, we always considered the initial
population $N_B(t=0)=N/2$. An alternative strategy could be to start
from the quasi-stationary state. However, when $N$ is small and/or
$L\leq L_c$, as seen in the previous section, the quasistationary
state is not so well defined from a physical point of view.

Let us first consider the extinction region ($L<L_c$). In this case,
for the continuous model, extinction is signaled by the density field
exponentially going to zero with rate given by the first eigenvalue,
$\lambda_1=1 - ({L_c}/{L})^2$. In particular, at sufficiently long
times we have that the biomass per unit length behaves as
$\mathcal{B}_C(t) \sim b \, e^{-|\lambda_1| t }$, with $b$ some
constant.  In the discrete model, it is natural to assume that
extinction will take place at a time $T_e$ such that
$\mathcal{B}_C(T_e) \sim 1/N$, implying, apart from a constant $b'$,
the logarithmic dependence on $N$
\begin{equation}
T_e \sim \frac{1}{|\lambda_1|} \log N + b',
\label{eq:Te_ext}
\end{equation}
which is consistent with the numerics, at least for $N$ large enough
(Fig.~\ref{fig:5}). We observe that the logarithmic behavior
(\ref{eq:Te_ext}) is also found in non-spatial models
\cite{lande1993}.

Conversely, for $L>L_c$, $T_e$ displays an
  exponential dependence on $N$ (Fig.~\ref{fig:5}), this is a quite
  robust feature observed in non-spatial models
  \cite{lande1993,nasell2001,dss2005} where it is derived either with
  the diffusive approximation  or more rigorous large-$N$
  approximations (see Ref.~\cite{ovaskainen2001} for a compact
  review).  Here, the spatial extension complicates  the use of those well
  established techniques. However, we observe that the exponential
  behavior for the time to extinction is consistent, at least for $L$
  and/or $N$ large enough, with the Gaussian behavior of the
  quasistationary distribution observed in the previous  section (see
  Fig.~\ref{fig:4}). Indeed, we can assume
\begin{equation}
P_{QS}(N_B) \propto \frac{1}{\sigma_N}\exp\left(-\frac{\left(N_B - \lra{N_B}\right)^2}{2 \sigma_N^2}\right) 
\label{eq:pNB}
\end{equation}
and it is reasonable to expect that the extinction time will be
controlled by the small $N_B$ tail of the quasistationary
distribution.  Roughly, the idea is that if before the extinction a well
defined quasistationary state  sets in, then the extinction will
take place when $N_B\approx 1$~\footnote{In other terms one assumes
  that once $N_B=1$ the extinction takes place with certainty.} and
the time to reach this state will be proportional to the inverse of
its probability so that $T_e \sim 1/P_{QS}(N_B \approx 1) \approx 1/
P_{QS}(N_B \to 0)$. Hence, using (\ref{eq:pNB}) we obtain
\begin{equation}
T_e \sim  \sigma_N \exp\left(\frac{\lra{N_B}^2}{2\sigma_N^2}\right) \sim
\sqrt{N} e^{a N}\,,
\label{eq:Te_pers}
\end{equation}
where we have used that $\langle N_B\rangle = N \mathcal{B}_{QS}$,
with $\mathcal{B}_{QS}$ independent of $N$ (as is true for large $N$, see
Fig.~\ref{fig:3}), and $\sigma_N =N\sigma_{\mathcal{B}_{QS}}\propto
\sqrt{N}$ as from the inset of Fig.~\ref{fig:3}. 

Numerical findings show that the logarithmic and exponential behaviors
are separated by the power-law $T_e \sim N^{\gamma}$ at the critical
patch size $L_c$ of the continuous model. In particular, our best fit
gives $\gamma=0.565$, which is not far from the analytical estimate
$T_e \sim N^{1/2}$ found in non-spatial stochastic logistic
models~\cite{nasell2001,dss2005}.  However, while not large, the
difference in the value of the exponent is nevertheless clearly
measurable and might be due to the spatial structure of the system
under study.

\begin{figure}[!h]
\centering
\includegraphics[scale=0.7]{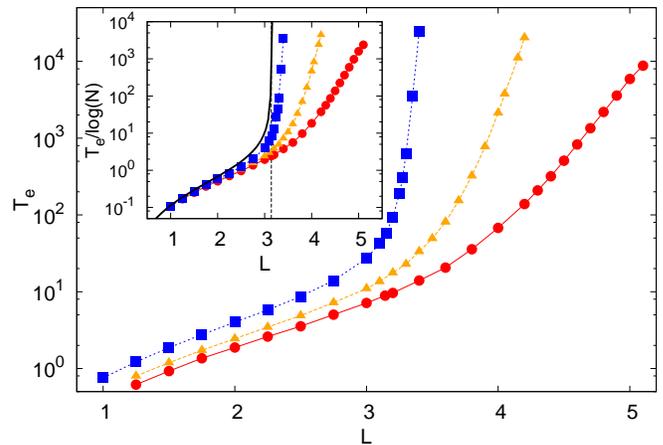}
\caption{\label{fig:6} (Color online) Mean extinction time $T_e$ as a
  function of the patch size $L$ for $N=40$ (red circles), $N=100$
  (orange triangles) and $N=1000$ (blue squares).  In the inset
  $T_e/\log N$ versus $L$ is compared to the asymptotic prediction
  (\ref{eq:Te_ext_L}) (solid black curve).
  The vertical dashed line marks the critical value $L_c=\pi$. The
  number of realizations used for the averages ranges from $500$ to
  $5000$ depending on the value of $N$.}
\end{figure}

We conclude briefly discussing the dependence of $T_e$ on $L$
(Fig.~\ref{fig:6}). When $L<L_c$, the extinction time $T_e$ is mainly
controlled by the eigenvalue $\lambda_1$ of the continuous KiSS model.
Indeed, from Eq.~(\ref{eq:Te_ext}), using the expression of
$\lambda_1$ we have
\begin{equation}
T_e \sim \frac{L^2}{L_c^2 - L^2} \log N\,,
\label{eq:Te_ext_L}
\end{equation} 
which is in fairly good agreement with the numerical data when $N$ is
large (inset of Fig.~\ref{fig:6}). 

\section{Conclusions}
\label{sec:concl}
We studied an individual-based reaction-diffusion system of ecological
interest. We focused on a model of a discrete population in a
favorable patch, surrounded by a deadly environment, which recovers
the KiSS model in the continuum limit. The presence of an absorbing
state together with demographic stochasticity makes population
extinction possible even when the continuous model would allow for
population survival. In such a case, however, the system attains a
quasi-stationary state that we have investigated at varying the system
and population sizes. We have shown that above the  KiSS critical patch size
the biomass in the quasi-stationary state recovers the continuum limit
value only if the population is large enough.  On the other hand, when
the population has a small number of individuals, the link between
biomasses, defined in the continuous and particle models, ceases to
exist. In particular, when the patch size tends to the critical
value the number of individuals required to recover the continuum
limit diverges.

Moreover, we have shown that the transition from extinction to survival
translates, in the discrete model, in the transition from a logarithmic
to an exponential dependence of the average time to extinction, $T_e$,
at varying the population size, $N$.  At the transition, these two
behaviors are separated by a power-law dependence $T_e \sim N^\gamma$
with an exponent definitively different from the analytical prediction
obtained in the non-spatial logistic model.    

We conclude mentioning that it would be interesting in the future to
investigate the effects of population discreteness in more complex
heterogeneous environments \cite{skt1986,kkts2003} and possibly in the
presence of advection \cite{dahmen2000life,rb2008,vibc2012}.  In such
cases, besides the survival/extinction transition, we expect that the
discreteness of the population will impact in nontrivial ways on the
spatial propagation properties of the population.

\appendix
\section{Effects of changes of the microscopic rules of the individual-based model}

In this Appendix we consider how two different modifications in the
individual-based model affect the system dynamics: the first one
regards variations of the interaction distance $R$ -- in order to
check the robustness of the particle model, the second one regards
particles' motility and consists in considering fixed A particles --
to investigate changes in the continuum limit.

\begin{center}
\begin{figure}[!hb]
\includegraphics[scale=0.7]{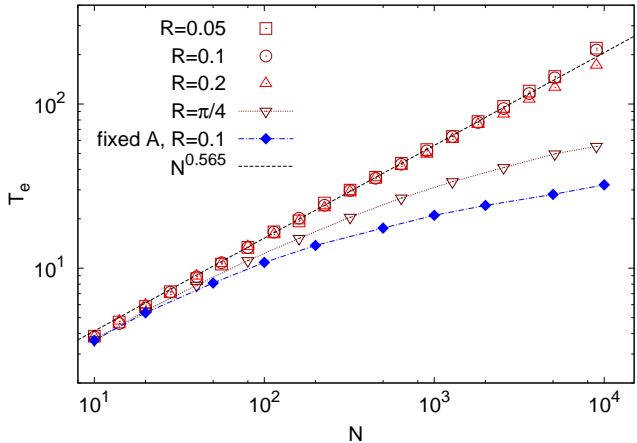}
\caption{ (Color online) Average extinction time vs $N$ for $L=L_c$ at varying $R$ as
  from the legend and considering non-motile A-particles. The power
  law behavior $T_e \sim N^{0.565}$, identifying the
  extinction/survival transition, is robust for variations of the
  interaction distance around the reference value $R=0.1$; for much
  larger values ($R=\pi/4$) the critical behavior is considerably
  altered.  The same happens if one considers a model in which
  particles of type $A$ do not move (filled diamonds).}
\label{fig:7}
\end{figure}
\end{center}

The interaction distance $R$ is a crucial parameter for the
individual-based model and it is associated to fluctuations of the
number of individuals. A significant variation of $R$ should thus have
important consequences, through its impact on the effective growth
rate.  In Figure~\ref{fig:7} we present a comparison between numerical
results obtained in the reference case $R=0.1$, used in this study,
and those obtained with different values of $R$. For illustrative
purposes only, we focus on the average extinction time as a function
 of $N$ by fixing $L$ at the critical value $L_c$.  As shown in the
figure, the power-law scaling $T_e \sim N^\gamma$ (with $\gamma \simeq
0.565$), and hence also the value of the critical patch size in the
discrete model, is robust if the interaction distance varies within
the same order of magnitude ($R=0.05$ and $R=0.2$).

The situation is different when $R$ becomes comparable with the patch
size: $T_e$ is no longer described by a power law (see the case
$R=\pi/4$ in Fig.~\ref{fig:7}) meaning that the discrete model is no
longer at the critical point. Indeed at large $N$ the curve
$T_e=T_e(N)$ bends towards a logarithmic shape as in
  Fig.~\ref{fig:5} when $L<L_c$.  Actually, a similar tendency
towards smaller values of $T_e$ characterizes the large $N$ behavior
for $R=0.2$, although the effect is so weak that it is difficult to be
seen in the figure.  The faster extinction is likely due to the fact
that increasing $R$ causes the hostile boundary conditions to be felt
deeper inside the favorable patch. In other words, while on average
the number of particles within $R$ gets larger, the fraction of $B$
particles does not, due to its depletion close to the boundary.

As for the modifications of the motility of particles, we study a case
in which $A$ particles are fixed, and the only other change in the
particle model is that, in order to preserve the total number of
individuals when a $B$ particle is absorbed at the boundary, a new $A$
particle is introduced at a random position (uniformly in the patch).
This model has a biological justification in the case in which
individuals of type $A$ (the nutrient) are non-motile (e.g. a plant
species) and individuals of type $B$ belong to a motile species
feeding on it. The new model shows a deep alteration in the critical
behavior at $L=L_c$ (see the filled diamond symbols in
Fig.~\ref{fig:7}) and the continuum limit of such a model is not given
by Eqs.~(\ref{eq:rd}-\ref{eq:logistic}) (see also Fig.~\ref{fig:1})
because particles of type $A$ and type $B$ do not undergo the same
diffusive process. One can expect that in this case the effective
growth rate is reduced due to less frequent encounters between
individuals of different types -- $A$ particles close to the
boundaries have less chances to turn into $B$ particles, and
extinction times result to be smaller with respect of the original
particle model.  This highlights once more that the relation between
  individual-based and continuous reaction-diffusion models is a
  delicate one, which needs to be carefully taken into account when
  modeling the evolution of biological populations.

%\bibliography{biblio}

\end{document}